\def\year{2019}\relax
\documentclass[letterpaper]{article} 
\usepackage{aaai19}  
\usepackage{times}  
\usepackage{helvet} 
\usepackage{courier}  
\usepackage[hyphens]{url}  
\usepackage{graphicx} 
\def\UrlFont{\rm}  
\usepackage{graphicx}  
\usepackage{multirow}
\usepackage{paralist}
\usepackage{epsfig}
\usepackage{booktabs}
\usepackage{amsmath}
\usepackage{amssymb}
\usepackage{xcolor}

\newcommand{\giedrius}{\textcolor{black}}
\newcommand{\ajay}{\textcolor{black}}
\newcommand{\yi}{\textcolor{black}}
\newcommand{\xhdr}[1]{\vspace{3pt}\noindent\textbf{#1}}

\title{Can You Explain That? Lucid Explanations Help \\Human-AI Collaborative Image Retrieval}
\author{
Arijit Ray, Yi Yao, Rakesh Kumar, Ajay Divakaran, Giedrius Burachas\\
SRI International, Princeton, NJ, USA
}
 
\begin{document}

\maketitle

\begin{abstract}
While there have been many proposals on making AI algorithms explainable, few have attempted to evaluate the impact of AI-generated explanations on human performance in conducting human-AI collaborative tasks. To bridge the gap, we propose a Twenty-Questions style collaborative image retrieval game, Explanation-assisted Guess Which (ExAG), as a method of evaluating the efficacy of explanations (visual evidence or textual justification) in the context of Visual Question Answering (VQA).
In our proposed ExAG, a human user needs to guess a secret image picked by the VQA agent by asking natural language questions to it. We show that overall, when AI explains its answers, users succeed more often in guessing the secret image correctly. Notably, a few correct explanations can readily improve human performance when VQA answers are mostly incorrect as compared to no-explanation games. Furthermore, we also show that while explanations rated as ``helpful'' significantly improve human performance, ``incorrect'' and ``unhelpful'' explanations can degrade performance as compared to no-explanation games.  
Our experiments, therefore, demonstrate that ExAG is an effective means to evaluate the efficacy of AI-generated explanation on a human-AI collaborative task.

\end{abstract}

\begin{figure}[ht!]
\centering
\includegraphics[height=185px]{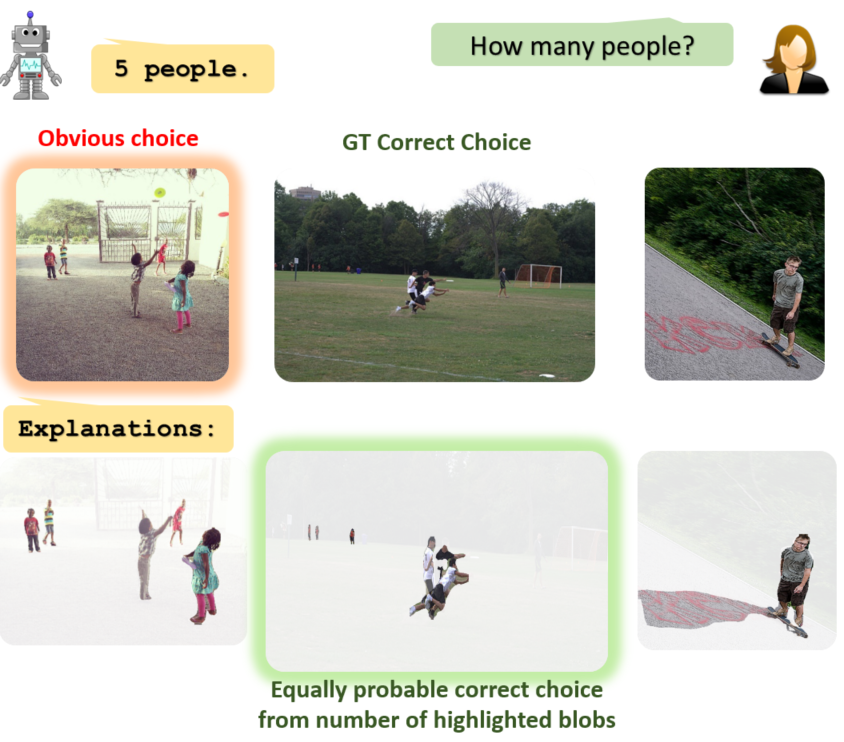}
\caption{We propose the ExAG game, where a human user needs to guess the secretly picked image by a VQA agent via asking natural language questions. The VQA agent answers ``5'' to the question ``How many people?''. Without any explanations, the user finds it difficult to judge the correctness of the agent's answer and hence, the first image seems an obvious choice as the secret image. The explanations (visualizations in bottom row) point out the critical evidence that the agent probably mistakenly sees five people in the second image as well. The user can take this into consideration while asking follow-up questions.}
\label{fig:teaserFig}
\end{figure}
\section{Introduction}

\ajay{Deep networks, as black-box models, often suffer from the lack of interpretability \cite{zhang2018visual}}. 
In the context of Visual Question Answering (VQA) \cite{antol2015vqa}, various methods for shedding light on the inner workings of these networks have been proposed --- pointing to evidence in the image and/or question \cite{lu2016hierarchical,kazemi2017show,xu2016ask,selvaraju2016grad} and human-interpretable text-based justifications \cite{park2018multimodal}, to mention a few. However, the empirical evidence that such explanations can actually be helpful for a human-machine collaborative task is lacking. 

To this end, we propose a Twenty-Questions \cite{lewis2008convention} style human-machine collaborative game, Explanation-Assisted GuessWhich (ExAG), using VQA \cite{antol2015vqa} as the backbone task to evaluate the efficacy of explanations. An explanation, in this context, refers to additional information output by the VQA that sheds light on the reasoning of the VQA agent for generating an answer given a question-image pair. For example, if the answer to a question ``what is in the image?'' is ``car'', an explanation could be pointing to salient components of cars such as the wheels and wind-shields. 
As shown in Figure \ref{fig:teaserFig}, in ExAG, human users and a VQA agent collaborate to retrieve a secret image, selected by the agent out of a set of visually similar images. 
The role of the agent is to help the human identify the secret image by answering questions asked by the human.  
Since the VQA is noisy in its answer predictions, finding the correct secret image requires humans to build a proper mental model of the VQA agent in order to decide which answer to trust. This makes ExAG a promising framework for evaluating the efficacy of explanations. 
Our hypothesis is that humans will succeed more often (i.e., winning rate) and quicker (i.e., using fewer questions) in finding the secret image when the machine explains its reasoning. 

We conducted two sets of ExAG games and collected user performance as a function of their usage of explanations. The first set (i.e., the at-will setting) allowed users to choose the use of explanations at will. This set of experiments provided preliminary evidence that human users spontaneously and increasingly prefer explanations even when their usage is penalized in the final score. 
The second set was conducted with a more controlled design to study the efficacy of each mode of explanations using a tighter metric (i.e., the controlled setting). 
We collected subjective ratings of explanation helpfulness perceived by the users before the outcome of the game was revealed (to avoid the influence of game win or loss on the perceived helpfulness).
We also independently collected subjective ratings of explanation correctness for better understanding of the relationship between explanation efficacy and quality. 
  

We show evidence that the ExAG game performance correlates to perceived explanation helpfulness and correctness ratings, making ExAG a suitable tool for evaluating explanation efficacy and quality. Helpfulness of an explanation is determined from users who rate how helpful they find the explanations while playing the game before knowing the outcome of the game. Correctness is determined by asking independent humans how relevant and correct the explanations are for the given image, question and answer. 
Interestingly, we also note that having a few ``correct'' explanations can help performance significantly in game rounds where answers are mostly incorrect.

Practical applications of our proposed ExAG can include image retrieval using free-form queries. For example, assisting disaster personnel, where a rescuer may have to rely on audio answers from a VQA machine because he/she is too busy to look at a video/image feed. It can also help medical professionals, where a doctor may use visual explanations to judge the confidence of a certain diagnosis among others.

\section{Related Work}

\xhdr{Explainable AI  }
Early work on explainable models involves template-based systems that spanned from medical systems \cite{shortliffe1984model} to educational settings \cite{lane2005explainable,van2004explainable}. 
Recent interest in explaining the inferences of a deep networks for computer vision applications includes introspective explanations that show the intermediate features of importance in making a decision \cite{lu2016hierarchical,park2018multimodal,xu2016ask,fong18net2vec,fong17interpretable,selvaraju2016grad,zeiler2014visualizing}, as well as post-hoc rationalization techniques such as justifying textual explanations \cite{park2018multimodal} and generating visual explanations \cite{hendricks2016generating}. We focus on using attention-based visualization of important image regions \cite{lu2016hierarchical,xu2016ask,kazemi2017show,teney2017tips}, object/scene detection \cite{szegedy2017inception,he2017mask} and semantically related question-answers to the asked question. It has been shown \cite{das2017human} that humans and machines look differently at images when answering questions. Hence, it is not obvious whether the above mentioned explanations are indeed helpful to humans. In this paper, we quantify how much explanations can help in human-machine collaboration performance.

\xhdr{Visual Question Answering }
We use VQA \cite{antol2015vqa} as the backbone task for our human-AI collaborative game. VQA is a vision-language task of answering natural language questions on images. Most of the effective approaches to VQA consist of works with attention on image features \cite{pythia18arxiv,lu2016hierarchical,teney2017tips,xu2016ask,kazemi2017show} guided by the question in order to answer it. 
We implement a comparable model that attends to both objects and free-form spatial regions in the image in a similar manner to \cite{pythia18arxiv}. 

\xhdr{20 Questions Game }
Our choice of the image-guessing game is a visual version of the popular 20-questions game, which is more formally, a specific version of the classic Lewis Signaling Game \cite{lewis2008convention}. 
There have also been efforts at training AI agents to play such an image-guessing game with humans/AI's \cite{DBLP:journals/corr/VriesSCPLC16} using reinforcement learning- \cite{das2017learning}. Such a game is used to evaluate the performance of visual conversational agents \cite{visdial_eval}. However, to the best of our knowledge, we are the first to use such a game to evaluate the effectiveness of explanations on human-machine collaborative tasks. 

\xhdr{Mental Model of an AI System } 
Along the lines of quantifying explanation efficacy, Chandrasekaran et al. quantified whether attention-based explanation improves human prediction of VQA performance \cite{chandrasekaran2017takes,chandrasekaran2018explanations}. While they show no significant increase in the ability to predict model outcome using attention-based explanations, we show that a combination of visual and textual explanations are helpful in a game setting where multiple rounds of question-answering are involved.  It is also shown that a combination of attention-based and textual explanations is helpful for predicting model performance \cite{park2018multimodal}. We use related question-answers as a form of textual explanation and also see similar trends for such a collaborative question-answering task. There have been works on evaluating the impact of visualizing model internals/workings on user trust, mental model understanding \cite{poursabzi2018manipulating} and performance \cite{nguyen2018believe}. While \cite{poursabzi2018manipulating} argued that displaying model internals might harm users ability to detect when model makes a mistake \cite{poursabzi2018manipulating}, we observe that in a collaborative setting with multiple rounds of interaction, a few correct explanations help improve performance on the task even when most of the model predictions in that session were incorrect.

\section{Game Outline}

In ExAG, there are two agents: a near state-of-the-art VQA deep learning model trained to answer questions about images (the ``VQA agent''), and a human volunteer (the ``player") who has to guess a secret image picked by the machine. A secret image is randomly picked from a pool of 1500 images. We select another $N-1$ images from the same pool using a difficulty measure based on the VGG16 \cite{simonyan2014very} FC7 distance. The difficulty level is adjusted so that the $N$ images are challenging enough to where it requires multiple rounds of VQA to identify the secret image. The player starts with $P_o$ points and is allowed to ask free-form questions to the VQA \yi{agent} in order to guess the \yi{secret} image. \yi{The final score is} $P=P_o-\sum_{i=1}^Q p_i$ if the correct image is guessed, where $Q$ is the number of questions asked and $p_i$ is the point deduction for each question. 
If the incorrect image is guessed, $P=0$. \yi{A success} is defined as the player correctly selecting the secret image while keeping $P>0$. \ajay{Players are encouraged to keep $P$ as high as possible.}

\subsection{The VQA Model}

\begin{figure}[h]
\centering
\includegraphics[height=155px]{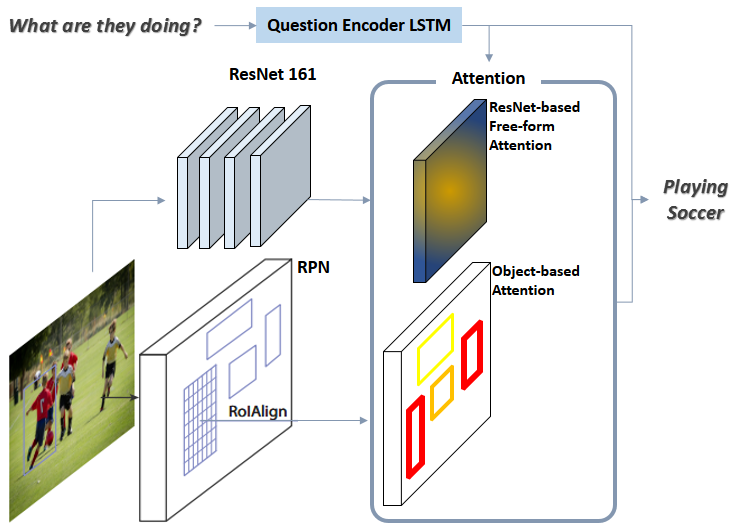}
\caption{Mask-RCNN-based VQA Model with attention on both global ResNet features (i.e., free-form attention) and region proposal network (RPN) features (i.e., object attention). This model is based on \cite{kazemi2017show} and \cite{teney2017tips}}
\label{fig:vqa_network2}
\end{figure}
 As shown in Figure \ref{fig:vqa_network2}, we used a near SOTA VQA model  
 that comprises both ResNet- \cite{szegedy2017inception} and Mask-RCNN-based \cite{he2017mask} image encoders
 and an attention mechanism to weigh the visual features depending on the question embedding. The question is encoded into an embedding using an LSTM. The weighted features are fed into a classifier that predicts an answer from 3000 candidates. 
 
\subsection{Modes of Explanations}
We define explanations as information given by the VQA agent that provides insight into why the VQA predicted a certain answer. The insight can be visualizations of the evidence used to infer the answer, such as weights applied to visual features (i.e., attention). It can also be rationalizations, such as stating the semantic beliefs about a fact that led to the answer. 
Below, we outline three modes of explanations and illustrate how they are used in ExAG.

\subsubsection*{Attention.}
\begin{figure}[t]
\centering
\includegraphics[height=79px]{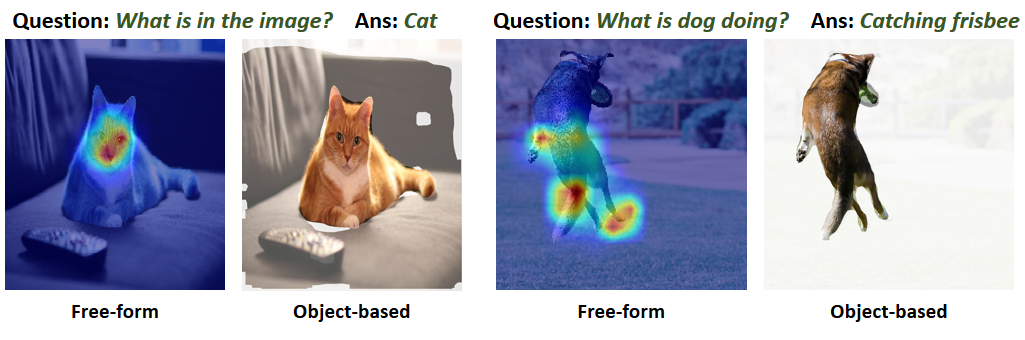}
\caption{Explanation mode based on attentions that highlight the relevant regions and objects in image to support the machine-generated answers for the given questions.}
\label{fig:attention2}
\end{figure}
\begin{figure}[t]
\centering
\includegraphics[height=60px]{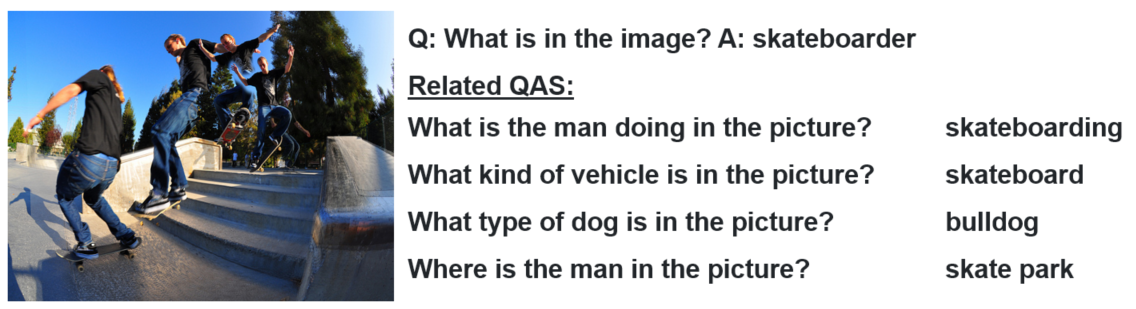}
\caption{Explanation mode based on related questions and answers. The given question is ``What is in the image?'' and the machine-generated answer is ``skateboarder''.}
\label{fig:relQAS_ex}
\end{figure}
We use attention masks computed based on the question asked to highlight spatial locations/objects in the image that are weighted more heavily in the inference process of answer prediction. 
We employ two types of attention layers - free-form attention that weighs visual features in the pixel-space and object-based attention that weighs object proposals in the image (Figure \ref{fig:attention2}).
A player can check if the attention masks correspond to the relevant part of the image given the question to determine if the machine generated answer is trustworthy.

\subsubsection*{Object and Scene Predictions (ObjScene).} We display a list of the most relevant object and scene predictions observed in the image. The relevance of an object/scene word $O$ is measured by $S(O)= \frac{dist(O,A)}{p(O|I)}$ where $I$ and $A$ denote the image and machine generated answer word, respectively, and $dist(O,A)$ denotes the Word2Vec distance \cite{mikolov2013distributed} between the object and answer words. $p(O|I)$ is calculated using the image encoder in VQA.  When object-based attention is used, this explanation mode is skipped since objects are already highlighted by attention masks. 
Under this explanation mode, the player needs to judge if the listed relevant objects are consistent with the visual contents in the image and machine-generated answer.  

\subsubsection*{Related questions and answers (RelQAs).} 

Five questions that are semantically close to the given question are retrieved from the VQA2.0 Validation Dataset. The closeness is measured via a semantic similarity based on the LSTM (from the question encoder in Figure \ref{fig:vqa_network2}) embedding distance over all the words in the pair of question and answer. Furthermore, questions with a high word overlap are rejected to avoid paraphrases (e.g., `what is in the image?' vs.~ `what is in the picture?'). The VQA agent generates answers to these related questions as part of the explanation. The player needs to judge the trustworthiness of machine-generated answer based on the correctness of and coherence among these pairs of related questions and answers (Figure \ref{fig:relQAS_ex}).

\begin{figure}
\centering
\includegraphics[height=80px]{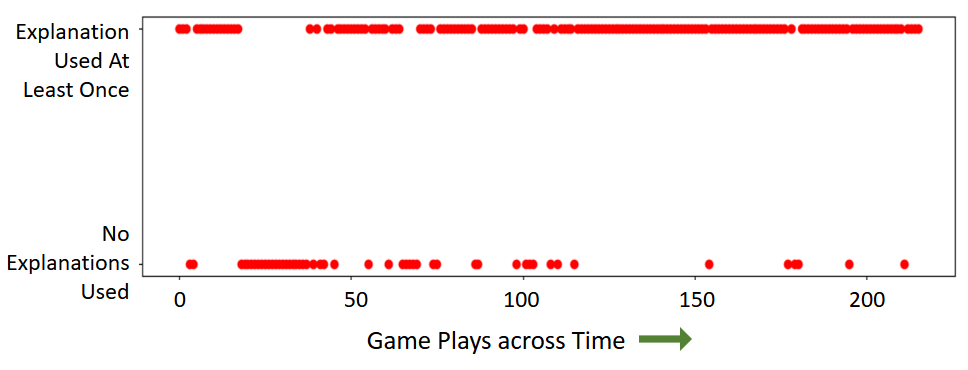}
\caption{When given a choice, players increasingly opt for explanations even when explanation usage is penalized - extra 2 points per question-answering round.}
\label{fig:explAdoption}
\end{figure}
\subsection{Game Settings}
\subsubsection*{At-will Setting.}
In this setting, $N=20$ images are selected and the player has the option of receiving explanations or not. Each question asked costs one point and explanation, if requested, costs an additional two points. All explanation modes are shown once the player chooses to receive explanations. This includes free-form attention for all images, ObjScene for the secret image, and RelQAs for the secret image. To be helpful of game wins, explanations need to be not only correct/coherent with the given answer but also sufficiently subtle/complete against distractor images.

\subsubsection*{Controlled Setting.}
In this setting, we show spatial attention, object-based attention and RelQAs for all the images. Since extra information (explanations) are given for all the images, the coherence between the explanation and given answer plays more important role in assisting game performance. In order to reduce the cognitive load on the players, we reduce $N$ from $N=20$ in the at-will setting to $N=5$ and make the images more similar to each other to maintain the same level of difficulty. Since we use object-based attention, the ObjScene explanation is not shown to avoid repetition. We randomly assign each explanation type to a group of AMT workers.
As users play the game, they are also asked to rate how helpful the explanations are after each round of question answering. 
Note that at the time of rating, players do not yet know the outcome of the game. So, their rating is not confounded by whether they succeeded or not, but likely reflects how helpful the explanations seemed in narrowing down the secret image and identifying the proper question to ask next.

\section{Experimental Conditions}

\begin{table*}[ht]
\centering
\caption{Human performances of ExAG in the controlled settings with different explanation modes. Overall, explanations improve game performance in terms of both win rate and averaged score. Explanations can help human players identify the secret image not only correctly but also with fewer questions (as reflected by higher scores).}
\label{table:ExplPerf}
\begin{tabular}{lcccccccccc}
\hline
\textbf{}               & \multicolumn{2}{c}{\textbf{With Expl}} & \multicolumn{2}{c}{\textbf{No Expl}} & \multicolumn{2}{c}{\textbf{Group Baseline}} & \multicolumn{2}{c}{\textbf{Overall Improv}} & \multicolumn{2}{c}{\textbf{Stat Sig}} \\ \cline{2-11} 
\textbf{}               & \textbf{Score}   & \textbf{Win Rate}   & \textbf{Score}  & \textbf{Win Rate}  & \textbf{Score}      & \textbf{Win Rate}     & \textbf{Score}      & \textbf{Win Rate}     & \textbf{p}         & \textbf{conf}    \\ \hline
\textbf{Attention}      & 6.23             & 66.67               & 6               & 64.92              & 5.66                & 62.1                  & 0.66                & 5.52                  & 0.1                & none             \\
\textbf{Rel QAs}        & \textbf{6.8}     & \textbf{71.48}      & \textbf{6.03}   & \textbf{64.54}     & \textbf{6.02}       & \textbf{65.45}        & \textbf{1.23}       & \textbf{10.33}        & \textbf{0.0019}    & \textbf{99\%}    \\
\textbf{Both}     & 6.44             & 69.03               & 5.83            & 63.25              & 5.68                & 63.75                 & 0.87                & 7.88                  & 0.03               & 90\%             \\
\textbf{Overall}        & 6.52    & 69.29      & 5.97   & 64.3      & 5.81       & 63.85        & 0.95       & 8.14         & 0.0015    & 99\%    \\
\textbf{Overall inc no} & \textbf{6.52}             & \textbf{69.29}               & \textbf{5.74}            & \textbf{61.85}              & \textbf{5.57}                & \textbf{61.15}                 & \textbf{0.95}                & \textbf{8.14}                  & \textbf{0.0015}             & \textbf{99\%}             \\ \hline
\end{tabular}
\end{table*}

\begin{figure}
\centering
\includegraphics[height=145px]{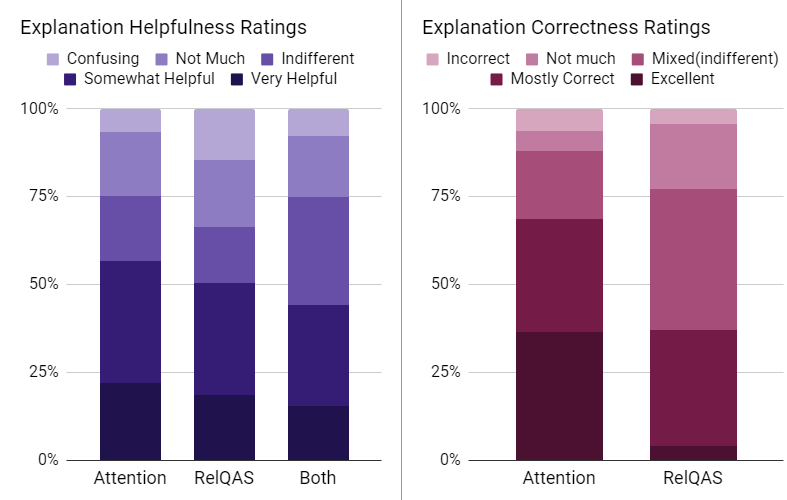}
\caption{Histogram of ratings of how ``helpful'' and ``correct'' explanations were while playing the game. ``Helpfulness'' ratings were given by workers while playing games before knowing the game outcome. ``Correctness'' was rated by three independent workers for explanations of the secret image.}
\label{fig:expl_help_ratings}
\end{figure}

\subsection{At-will Setting}

\begin{figure*}[ht]
\centering
\includegraphics[height=335px]{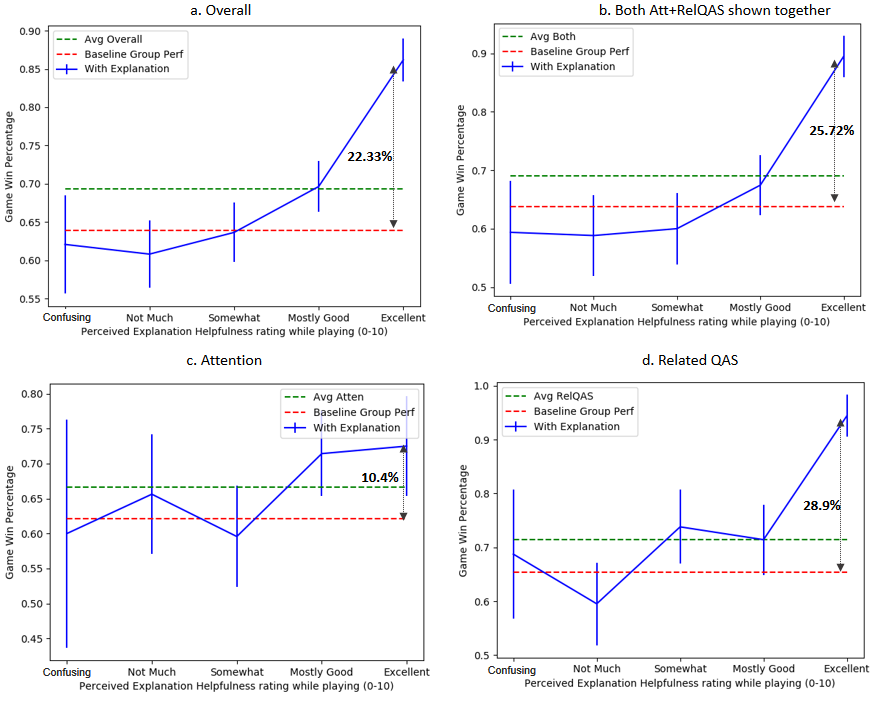}
\caption{\giedrius{Game win percentage as a function of user perceived helpfulness rating.} Baseline is first 5 no-explanation performance for the same group of workers. Helpfulness was self rated by workers before they knew the GT image or the outcome of the game. We see that overall, \textit{excellent} explanations significantly improve performance.}
\label{fig:expl_help_gamewinrate}
\end{figure*}
\begin{figure}[ht!]
\centering
\includegraphics[height=105px]{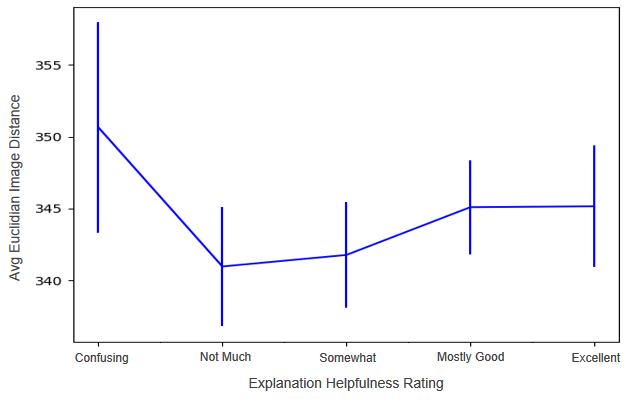}
\caption{Image sets were of similar difficulty (as measured by deep image feature distance of distractor images to GT image) for explanations that were rated ``Excellent'' as compared to ``Confusing'', thus not confounding improve in performance for ``Excellent'' explanations.}
\label{fig:diffculty_level}
\end{figure}
In this setting, the ExAG game was played in a competitive setting (with cash rewards for the team that won the most within a stipulated time) by about 60 people grouped into 6 teams. The players were free to choose explanations or forgo them. Using explanations resulted in an additional 2 point penalty and time loss since explanation generation takes more time than just answer generation.   

Of 206 total games played, the average win rate was $43\%$. We divided the game plays into games where explanations were never used ($N=49$) and those where explanations were used at least once ($N=157$). The win rate with  explanations was $47\%$ and $28\%$  when explanations were never used. The z-test for proportions indicates that this is a statistically significant difference at 95\% confidence level (p=0.019).

Moreover, as the players proceeded to play the ExAG game, they tended to opt for using explanations even though that resulted in additional 2 point penalty and time loss. Figure \ref{fig:explAdoption} shows this spontaneous adoption of explanation with increasing number of plays. A z-test comparing the proportion of games using the explanations during the first half of plays ($61.2\%$, $N=103$) vs. during the second half ($91.3\%$, $N=103$) indicates highly statistically significant increase in explanation utilization ($p<10^{-5}$).

\subsection{Controlled Setting}

\begin{figure*}[t]
\centering
\includegraphics[height=311px]{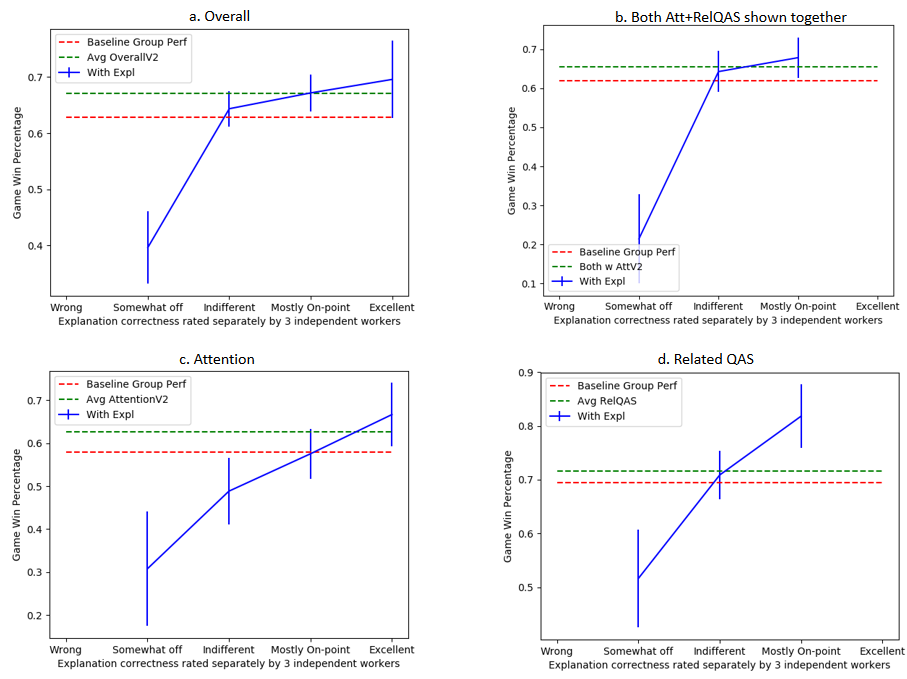}
\caption{\giedrius{Game win percentage as a function of independently rated explanation accuracy.} Baseline performance was adjusted to reflect the baseline of only that subset of games where correctness ratings were collected. Overall, we see that incorrect explanations hurt performance, while correct explanations are not sufficient to improve performance compared to no-explanation games.}
\label{fig:expl_corr_gamewinrate}
\end{figure*}

\begin{figure}[ht!]
\centering
\includegraphics[height=167px]{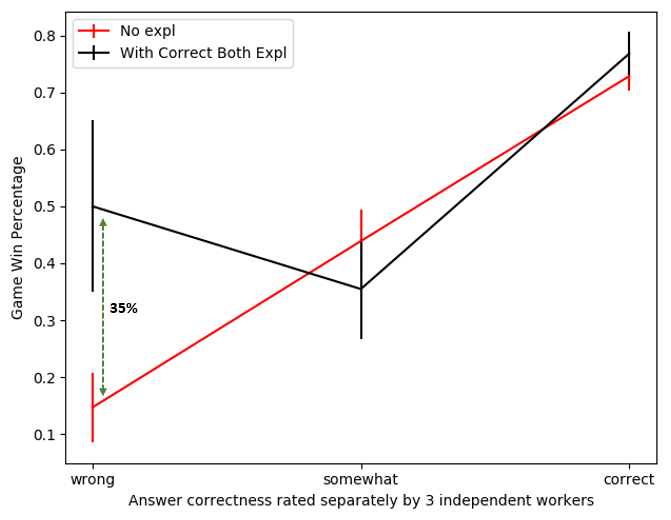}
\caption{\giedrius{Explanations help when VQA answers are wrong. Without explanations (red line), if the answer from the VQA is wrong, user performance drops dramatically. However, at least a few good explanations (black line) help reveal VQA answer correctness so that it can be taken into account.} Hence, game performance without explanations is much lower when answers are wrong than with explanations. The black line is defined as games where there was at least one explanation above \textit{indifferent}. }
\label{fig:both_explAnsWrongHelp_gamewinrate}
\end{figure}

This setting shows explanations for all images and hence, players have to rely on the coherence of explanations to aid their game (i.e., if explanations are good, the secret image's explanation will be more coherent with answer shown compared to the explanations for the distractor images). All analyses henceforth are performed on this setting. 

The games were played by $69$ individual AMT workers.
The instructions were to try to win the game by guessing the secret image correctly and were warned that a lack of effort (e.g., randomly select answers) would lead to rejection. 
For AMT worker selection, location was set to the US only to recruit workers with proper English skills, and worker qualification threshold was set to above $98\%$ (number of HITs~$>1000$) to ensure the quality of game plays. The workers played 1469 games in total covering four explanation modes - attention, RelQAs, both shown together (referred to as `Both'), and also without any explanations (``WithoutExpl'' group). 

Each individual worker always saw only one mode of explanation while playing. Each worker was instructed to play at least four sessions with each session consisting of five games. Sessions with and without explanations were alternated. For instance, the first session does not provide explanations, followed by a session with assigned choice of explanations. The first block is used as the ``group baseline'' no-explanation performance for the assigned explanation worker group. When reporting ``no expl'' performance  for a group, we average the first and third blocks and ``expl'' performance is the average of the second and fourth blocks. 

Note that we also studied the performance of each round of the ``WithoutExpl'' control group (i.e., group that never saw any explanations) to examine whether the performance of later rounds would be improved given the experiences of earlier rounds. No statistically solid improvement was observed mostly because only five consecutive tries were allowed and image sets with sufficient variety were used in each trial. Based on this study, we could directly compare performances in each round without the need to normalize scores to balance learning effect that can potentially confound the player's performance. 


\section{Results}
\subsubsection*{\giedrius{Overall, explanation usage improves game performance:}}
\giedrius{The overall impact of explanations on the performance of ExAG is summarized in Table \ref{table:ExplPerf}. The `Overall' row averages across all explanation modes (i.e., the three rows above) whereas the `Overall inc no' also includes the performance of workers who never saw any explanations (i.e., the `WithoutExpl' group). For each of the modes, `With Expl.'and `No Expl.' list the performance with and without explanations, respectively. `Group baseline' is the baseline performance without explanations (the first 5 rounds) of that worker group. `Overall Improv' shows the improvement with respect to `Overall inc no'. The game score starts from 10 ($P_o=10$) with each question asked costing one point ($p_i=1$). We observe that the game win rate is statistically significantly ($p=0.0015$, $99\%$ confidence) improved by explanations overall and fewer questions were required to guess correctly (as suggested by higher scores).`RelQAs' improves game performance the most ($p=0.0019$, $99\%$ confidence), followed by `Both' ($p=0.03$, $90\%$ confidence).} We don't see a statistically significant effect of attention-based explanations when presented in isolation, which is consistent with observations from prior works \cite{chandrasekaran2017takes,chandrasekaran2018explanations}. 

We also analyze how the difficulty level of the selected image set affects the game performance. We use the $L_2$ norm on the VGG16 FC7 feature \cite{simonyan2014very} to evaluate the distance between each images and the secret image and the averaged distance as the overall difficulty level of the set- lower distance meaning higher difficulty. We observe that for game rounds with explanations, the average difficulty level is similar between winning and losing game rounds. However, for game rounds without explanations, the difficulty level of winning games is much lower than that of the losing games. 
This may suggest that explanations help players to identify critical cues about how the VQA answers questions which may not be visually salient otherwise.   

To understand the effect of explanation quality on performance, we collect two types of ratings for the explanations 1) While playing the game, workers are asked to self rate their ``perceived helpfulness'' of the explanation for zeroing down on an image after each question asked. 
2) We separately collect the correctness of the explanations and answers from 3 independent AMT workers by showing them the answer and explanation for the asked question and the secret image. 
Below, we use these ratings to analyze how ``perceived helpfulness'' and ``correctness'' of explanations and answers affect game performance.

\subsubsection*{\giedrius{User-Rated helpful explanations improve game performance the most:}}

We collect self-reported helpfulness ratings of explanations as workers receive answers and explanations for their questions (Figure \ref{fig:expl_help_ratings}). We analyze game performance as a function of these ratings to see if explanations perceived as ``helpful'' do help game performance. Since workers didn't know the secret image or the game outcome while rating, their decision was not affected by the (future) play outcome and likely reflected the helpfulness of explanations in narrowing down their image choices. The workers were asked to rate the explanations according to the following options: ``Helping a lot'' (refered to as \textit{Excellent} here-on), ``Mostly helping'' (\textit{Mostly Good}), ``Somewhat Helpful'' (\textit{Somewhat}), ``Not helping much'' (\textit{Not Much}) and ``Completely Confusing'' (\textit{Confusing}). The histogram of the ratings across all the games is shown in Figure \ref{fig:expl_help_ratings} and the explanations are mostly perceived as helpful. 

We see that overall (Figure \ref{fig:expl_help_gamewinrate}a), explanations perceived as \textit{Excellent} significantly increase game performance. When explanations are rated less helpful, performance is similar to playing without explanations supporting the notion that workers seem to ignore the explanations, or somewhat degraded as the workers were confused by them.
Notably, we calculate the average difficulty levels of the candidate image sets for each game round and observe no confound of image sets being more easy when explanations were rated more helpful (Figure \ref{fig:diffculty_level}).

Next we break down explanation helpfulness-dependent performance by explanation modes. Figure \ref{fig:expl_help_gamewinrate}b
shows that combining attention and RelQAs improves performance significantly for explanations rated as \textit{Excellent}, but hurts performance slightly when rated below \textit{Somewhat}. Figure \ref{fig:expl_help_gamewinrate}c indicates that attention-based explanations don't help much on their own, however, when they seem very helpful for making a choice, they do help game performance slightly. \textit{Excellent} RelQAs were the most helpful for game performance when presented in isolation as in Figure \ref{fig:expl_help_gamewinrate}d. 
We reason that this is probably due to the consistency of related question answers being a slightly better indicator of VQA answer accuracy. We calculate the correlation coefficient of the explanation correctness to answer correctness and observe that it is $0.37$ for related question-answers as compared to $0.33$ for attention. This combined with the ease of parsing textual attention than heatmaps probably make related QA explanations more effective. 

\begin{figure*}[ht!]
\centering
\includegraphics[height=295px]{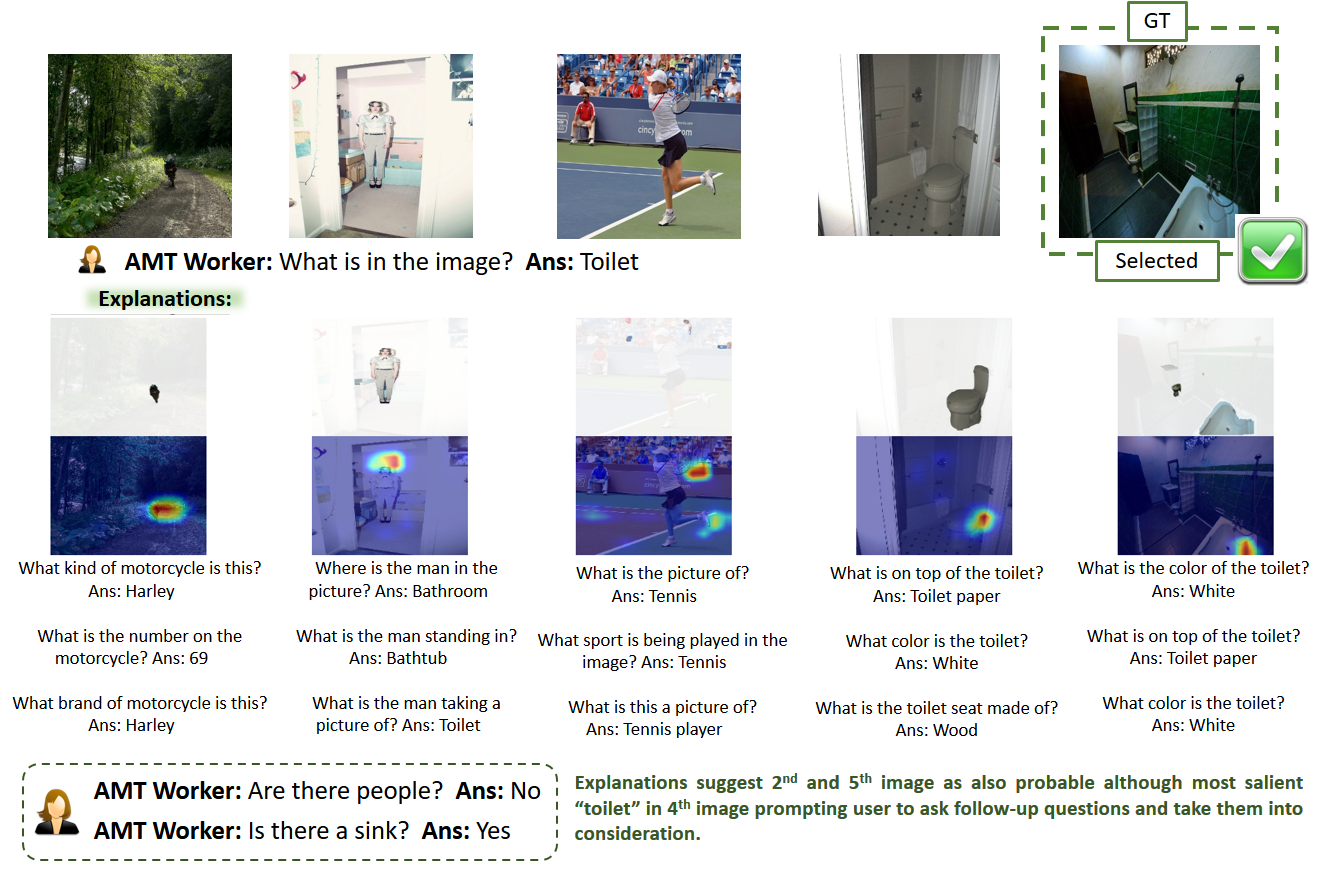}
\caption{An ExAG game round where explanations help the user in winning the game. Even though the most prominent `toilet' is in the fourth image, explanations make it clear to the users that the fifth image could also be the secret image. Explanation suggest that the machine probably mis-detected the `bathtub' or `sink' for a `toilet'. This hints the user to ask follow-up questions like ``is there a sink?" to finally select the correct image.}
\label{fig:WithExplPlay}
\end{figure*}

\subsubsection*{\giedrius{Independently-rated incorrect explanations degrade game performance:}}

To analyze by explanation correctness, we ran a separate AMT task to collect correctness ratings of the explanations for the secret image, the question that was asked and the answer given by the VQA model. Three independent workers were asked to rate the explanations for the given question and image according to the following options: ``Exactly on-point'' (referred to as \textit{Excellent} hereon), ``Mostly on-point'' (\textit{Mostly on-point}), ``Indifferent'' (\textit{Indifferent}), ``Somewhat off'' (\textit{Somewhat off}) and ``Completely Wrong'' (\textit{Wrong}). The histogram of ratings is shown in Figure \ref{fig:expl_help_ratings}. 

Figure \ref{fig:expl_corr_gamewinrate} shows game performance as a function of explanation correctness ratings. We see that for modes with overall (\ref{fig:expl_corr_gamewinrate}a), combined (\ref{fig:expl_corr_gamewinrate}b) and attention-based (\ref{fig:expl_corr_gamewinrate}c), correct explanations are not sufficient for improving game performance, while, incorrect explanations can severely degrade the performance. This suggests that giving incorrect explanations can make players disbelieve a correct answer and hence fail games. 
We observe that while RelQAs has very few \textit{Excellent} cases, as long as they are \textit{Mostly on-point}, they help game performance substantially (\ref{fig:expl_corr_gamewinrate}d). 

\subsubsection*{\giedrius{A few correct explanations help when VQA answers are mostly incorrect:}}

As noted before, we observed that correct explanations do not necessarily help game performance. Further examination indicates that game performance is indeed influenced by the combined effect of the correctness of explanations and machine-generated answers. We collect the answer correctness ratings for the ExAG games through an independent AMT experiment. The workers were required to rate the answer for a given question and image pair as either ``correct'', ``somewhat correct'' or ``completely wrong''. We analyzed the game performance with varying average VQA answer correctness in game rounds with at least one correct (above \textit{indifferent}) explanation and without. 

As displayed in Figure \ref{fig:both_explAnsWrongHelp_gamewinrate}, we see that having at least a few correct explanations (black line), interestingly, helps user performance significantly ($p=0.013$, $95\%$ confidence) in games where the VQA answers are mostly wrong as compared to no-explanation games (red line). 
This suggests that having explanations helped workers in identifying potentially incorrect answers which motivates them to ask clarification questions, a few of which could have had correct answers with correct explanations that the users could pick out. 
\giedrius{As expected, without explanations (red line of Figure \ref{fig:both_explAnsWrongHelp_gamewinrate}), game performance degraded as VQA answers got less accurate, suggesting that a player had no way of telling if an answer was correct or not without explanations.}

An qualitative run of ExAG is shown in Figure \ref{fig:WithExplPlay}. RelQAs explanations suggest that the VQA also understandably sees a `toilet' in the fifth image. This prevents the user from selecting the obvious fourth image choice straight away and prompts him/her to further ask questions like ``is there sink?'', eventually resulting in him/her selecting the correct  secret image. 

\section{Conclusion}

\giedrius{We propose the ExAG game as an evaluation framework for explanations in VQA.}
Our experiments provide empirical evidence that overall, explanations help improve performance on such a human-machine collaborative image guessing task.
When analyzed by user self-rated ``helpfulness'' and independently-rated ``correctness'',  helpful explanations (rated as \textit{excellent}) significantly improve performance, while \textit{incorrect} explanations degrade performance.
\giedrius{Moreover, since the self-rated helpfulness is not influenced by the outcome of the game, this suggests that users can use their insight into explanation helpfulness to decide when to trust and include them in decision making process for choosing the image.}

We also note that ``correct'' explanations, interestingly, help significantly when machine predictions are noisy (Figure \ref{fig:both_explAnsWrongHelp_gamewinrate}). Without explanations, users blindly trust incorrect machine predictions, which hurts game performance. With explanations, users ask follow-up questions and are able to succeed based on few correct explanations.

We believe that our ExAG framework can help in designing more accurate and helpful explanations that improve human-AI collaboration performance.

\section*{Acknowledgments}
This research was developed with funding from DARPA under the XAI program. The views, opinions and/or findings expressed are those of the authors' and should not be interpreted as representing the official views or policies of the Department of Defense or the U.S. Government.

\begin{quote}
\begin{small}
\bibliographystyle{aaai}
\bibliography{egbib}
\end{small}
\end{quote}

\section*{Appendix}

\subsection*{VQA Architecture Details}
\begin{figure}[ht!]
\centering
\includegraphics[height=145px]{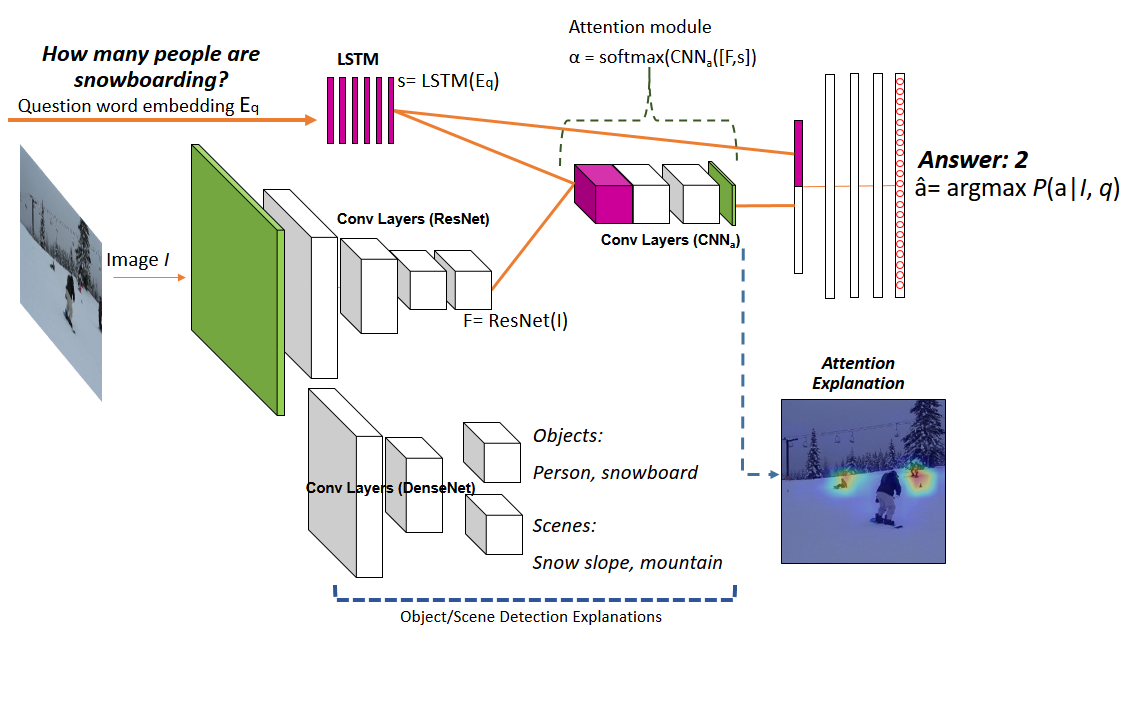}
\caption{ResNet-based VQA model used in the At-will game setting. This network attends to ResNet 161 features. This model is based on \cite{kazemi2017show}.}
\label{fig:vqa_network1}
\end{figure}

\begin{figure*}
\centering
\includegraphics[height=510px]{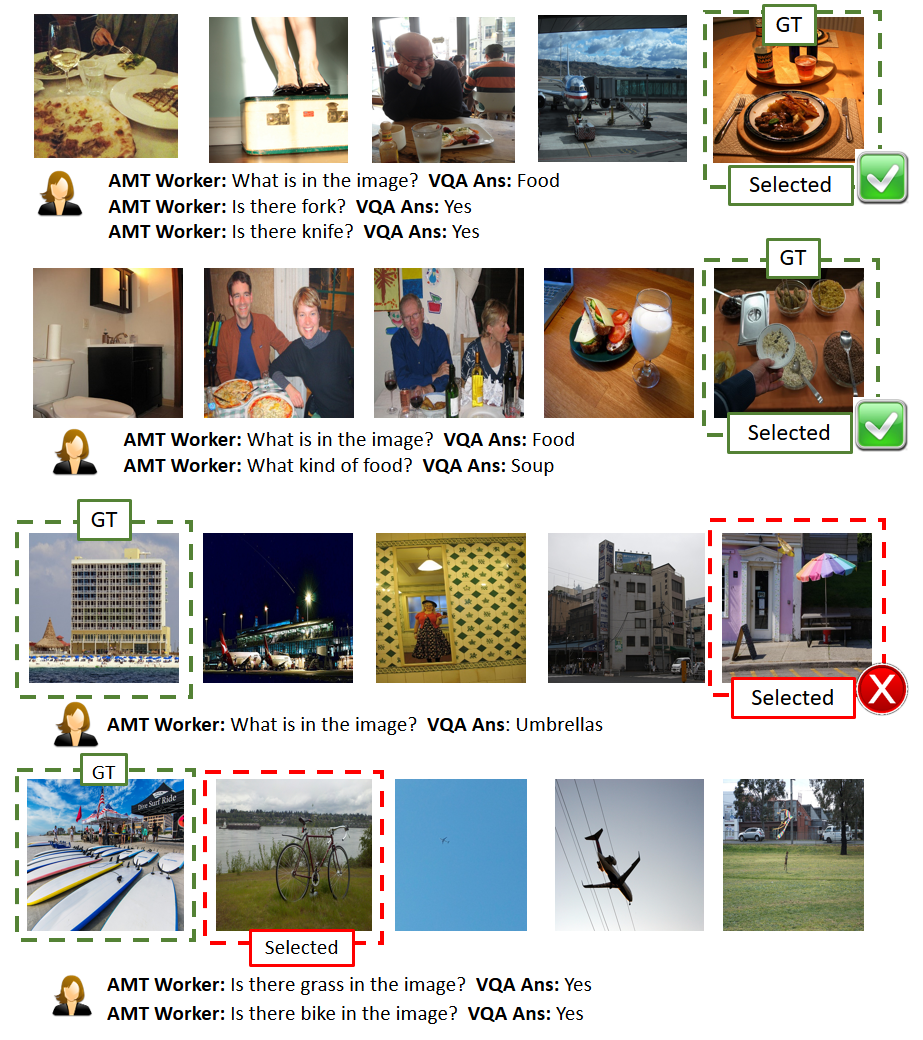}
\caption{This image shows game plays without explanations (each row is a game-play example). When the VQA is fairly accurate, a user is easily able to pick out the correct image as shown in row 1 and 2. However, when the VQA answer is incorrect, it leads to a reasonably wrong selection by the user since the user has to blindly trust the VQA answer.}
\label{fig:noExplPlay}
\end{figure*}

\begin{figure*}
\centering
\includegraphics[height=460px]{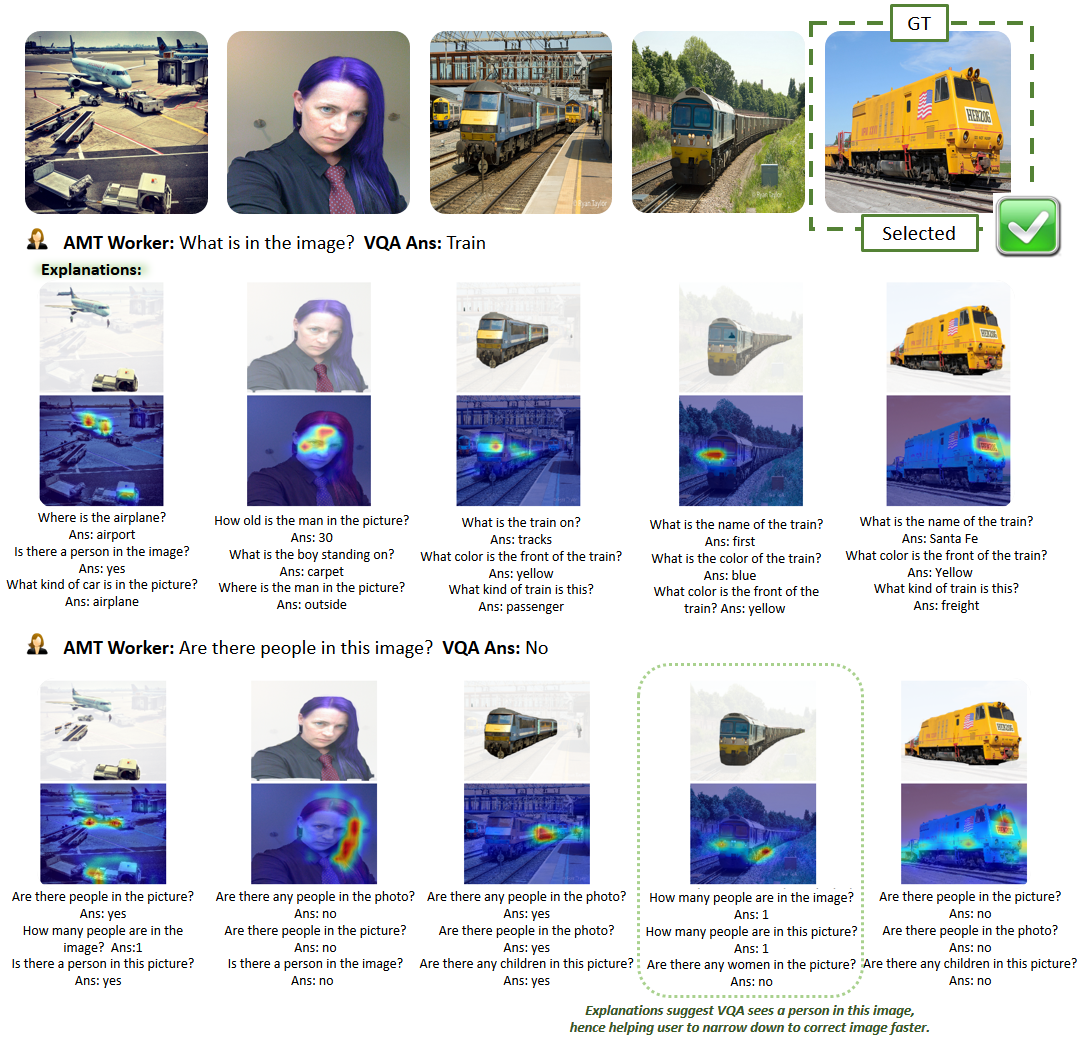}
\caption{This image shows a game with both explanations. Note how the explanations help confirm that the VQA indeed notices the somewhat hidden person inside the train in the fourth image from the left. This makes the user take that into consideration while guessing the secret image.}
\label{fig:WithExplPlay}
\end{figure*}

We use Tensorflow \cite{tensorflow2015-whitepaper} for all our implementations.
The network we use for At-will game setting is shown in Figure \ref{fig:vqa_network1}. Outlined below are the network details:
\begin{itemize}
\item Input Image - size $448 \times 448$. We center crop all images to the mentioned size during training and reshape during evaluation. We encode the image using a ResNet 161 \cite{szegedy2017inception} network. 
\item Question Input - Each question word is encoded using the Glove \cite{pennington2014glove} 300 dimensional embeddings before feeding into an LSTM word by word. We take the final 512 dimensional LSTM features as the question feature. Embeddings are fine-tuned.
\item Attention- For Resnet feature attention, We tile the question features (512) into $14 \times 14 \times 512$ and concatenate with $14 \times 14 \times 2048$ image features. Attention predicts a set of weights in the shape of $14 \times 14$ using a 2-D convolutional layer.
\item Answer classifier-  We concatenate weighted flattened ResNet features and question features and pass it through a fully connected layer to get 3000 answer logits. We compute a softmax to get probabilities.
\end{itemize}

\noindent For Controlled Game Setting, we use a VQA network as shown in Figure 2 in main paper. Details are as follows:
\begin{itemize}
\item Input Image - For ResNet161 \cite{szegedy2017inception} image features, we center crop images to $448 \times 448$ similar to Network 1 to get $14 \times 14 \times 2048$ dimensional features. We also use the Region Proposal Network from Mask RCNN \cite{he2017mask} to generate 100 object proposals per image. The input image size to Mask RCNN is $1024 \times 1024$ and the images are re-sized without cropping. We pool 1024 dimensional features from each of the 100 proposal boxes.  
\item Question Input - Each question word is encoded using the Glove \cite{pennington2014glove} 300 dimensional embeddings before feeding into an LSTM similar to network used in the at-will setting. The embeddings are fine-tuned. 
\item Attention- We have two attention modules - one for attending to the ResNet features (same as Network 1) and one for attending to the 100 object proposals in the image (object attention). Resnet feature attention is same as at-will setting. For object proposal attention, we concatenate the question features (512) to each 1024 image feature for the 100 proposals. Attention predicts a weight of shape $100 \times 1024$ using a 1-D convolutional layer.  
\item Answer classifier- We concatenate weighted flattened ResNet features, averaged weighted object features and question features and pass it through a fully connected layer to get 3000 answer logits. We compute a softmax to get probabilities.
\end{itemize}

\subsection*{Qualitative Results}
Figure \ref{fig:noExplPlay} and Figure \ref{fig:WithExplPlay} show qualitative examples of plays without and with explanations and how explanations may help in choosing the correct image more often when VQA answers are noisy.

\end{document}